\documentclass[aps,pre,twocolumn,a4paper]{revtex4}
\usepackage{graphicx}
\usepackage[dvips]{color}

\begin{document}

\title{On the efficiency of path sampling methods for the calculation of free energies from non-equilibrium simulations}

\author{Wolfgang Lechner and Christoph Dellago}
\affiliation{Faculty of Physics, University of
Vienna, Boltzmanngasse 5, 1090 Vienna, Austria}

\date{\today}

\begin{abstract}
According to the Jarzynski theorem, equilibrium free energy differences can be calculated from the statistics of work carried out during non-equilibrium transformations. Although exact, this approach can be plagued by large statistical errors, particularly for systems driven strongly away from equilibrium.  Recently, several approaches have been  suggested to reduce these errors. In this paper we study the efficiency of these methods using two models for which analytical solutions exist. 
\end{abstract}

\maketitle

\section{Introduction}

The calculation of free energies is central to many applications of computer simulations ranging from the calculation of ligand affinities to the study of phase equilibria in condensed materials. Since the free energy is a quantity related to the phase space volume available to a system, its calculation is non-trivial in most interesting cases and in the past decades considerable effort has been devoted to developing efficient free energy computation techniques \cite{FRENKEL_SMIT,FREE_ENERGY_BOOK}. 

Recently, Jarzynski has shown how free energies can be determined from non-equilibrium simulations \cite{jarz}. In this so called {\em fast switching} or {\em fast growth} method a control parameter coupled to the Hamiltonian of the system is switched from an initial to a final value at a finite rate. By changing the external parameter, the work $W$ is performed on the system. As a consequence of the second law of thermodynamics the average of this work is larger than the free energy difference $\Delta F$ between the equilibrium states corresponding to the final and initial value of the control parameter: 
\begin{equation}
\langle W\rangle  \geq \Delta F.
\end{equation}
Here, the angular brackets denote an average over pathways starting from the equilibrium state corresponding to the initial value of the control parameter.  The equal sign in this inequality, known as {\em Clausius inequality} or {\em maximum work theorem}, holds only if the external parameter is switched reversibly.

In 1997 Jarzynski demonstrated that the Clausius inequality can be turned  into an equality by averaging the work exponential rather than the work itself \cite{jarz,JARZ_PRE_97}:
\begin{equation}
\label{equ:jarzynski}
\langle e^{-\beta W}\rangle=e^{-\beta \Delta F},
\end{equation}
where $\beta = 1 / k_{\rm B}T$ reciprocal temperature. The Jarzynski equation (\ref{equ:jarzynski}) has been proven for several types of dynamics ranging from Newtonian and thermostatted dynamics to Langevin and Monte Carlo dynamics \cite{jarz,JARZ_PRE_97,gavin_jstatphys,Hummer_Szabo,Evans,Dellago_Paschinger}. Essentially, the theorem is valid if the dynamics preserves the canonical distribution for fixed control parameter \cite{JARZ_PRE_97,Dellago_Paschinger}.  

Remarkably, the validity of the Jarzynski equality does not depend on how the control parameter is switched from its initial to its final value. (In particular, it is not even necessary to maintain a fixed protocol throughout the simulation or the experiment as long as the initial and final value of the control parameter are fixed.) Thus, the switching process can be carried out at arbitrary speed. For infinitely slow switching, Jarzynski's method reduces to Kirkwood's coupling parameter method, or  thermodynamic integration \cite{Kirkwood}, while for infinitely fast switching one obtains Zwanzig's thermodynamic perturbation method \cite{Zwanzig}. Thus, these two well known free energy calculation methods can be viewed as limiting cases of the fast switching procedure.

Whether the freedom derived from the arbitrariness of the switching protocol can be used to develop more efficient free energy calculation methods is an interesting and practical question. A recent analysis indicates that the statistical errors encountered in evaluating exponential averages such as that of Eq. (\ref{equ:jarzynski}) prevent fast switching methods from exceeding the efficiency of conventional methods such as umbrella sampling or thermodynamic integration \cite{jarTPS}. This conclusion seems to remain true even if path sampling techniques are used to focus on those non-equilibrium pathways that yield the largest contribution to Jarzynski's exponential average \cite{SUN,YTREBERG}. Note that these path sampling approaches can also be combined with the large time step method for fast switching simulations \cite{LECHNER}. 

In this paper we analyze the efficiency of several fast switching methods including the path sampling approaches mentioned above for two analytically solvable models: a particle dragged through a viscous fluid in a harmonic trap \cite{MAZONKA_JARZYNSKI} and an ideal gas in an expanding piston \cite{GROSBERG_LUA_PEDAGOGICAL}. For these two models the work distributions can be calculated analytically for arbitrary switching speeds. Using these work distributions, we determine the statistical errors and, from them, the computational efficiency that one would obtain in fast switching simulations of these systems. The results presented in this paper confirm our previous conclusions based on actual simulations \cite{jarTPS}. Namely, fast switching simulations do not outperform umbrella sampling and thermodynamic integration in most interesting cases.    

The remainder of the paper is organized as follows. Simulation methods are described in Sec. \ref{sec:methods}. Expressions for error and efficiency estimates are presented in Secs. \ref{sec:error} and \ref{sec:efficiency}. In Sec. \ref{sec:results}  these expressions are then applied to the two model systems. Conclusionss are provided in Sec. \ref{sec:conclusion}.

\section{Fast switching methods}
\label{sec:methods}

In this section we review the fast switching methods whose efficiency we will analyze in later sections. To set the notation, consider a system with a Hamiltonian ${\cal H} (x, \lambda)$ that is a function of the state of the system $x$ and the external control parameter $\lambda$. Depending on the system, $x$ may include positions and momenta of all particles or the positions only. The free energy of the system at temperature $T$ and fixed control parameter $\lambda$ is given by
\begin{equation}
F_\lambda(\beta) = - k_{\rm B}T \ln Q_\lambda(\beta),
\end{equation}
where the normalizing factor
\begin{equation}
Q_\lambda(\beta) = \int dx\; e^{-\beta {\cal H}(x, \lambda)}
\label{equ:partition}
\end{equation}
is the partition function of the system (up to a constant factor). The free energy difference between two states $A$ and $B$  corresponding to two different values of the control parameter, $\lambda_A$ and $\lambda_B$, respectively, is given by
\begin{equation}
\Delta F = F_{\lambda_B} - F_{\lambda_A}=-k_{\rm B}T \ln \frac{Q_{\lambda_B}}{Q_{\lambda_A}}.
\end{equation}
State $A$ is described by the Hamiltonian ${\cal H} (x, \lambda_A)$ and state $B$ by  ${\cal H} (x, \lambda_B)$. We now imagine that the control parameter $\lambda$ is continuously switched from the initial value $\lambda_A$ to the final value $\lambda_B$ according to some protocol $\lambda(t)$ within a total time $\tau$. As the control parameter changes, the system evolves in time tracing out a trajectory $x(t)$. Here, $x(t)$ denotes a complete trajectory describing the system from time $t=0$ to time $t=\tau$. The change in energy that is due to the changing control parameter is the work $W$ carried out on the system along the particular trajectory $x(t)$:
\begin{equation}
W[x(t), \lambda(t)]=\int^\tau_0 dt\; \dot \lambda(t) \left. \frac{\partial {\cal H}}{\partial \lambda}\right|_{x=x(t)},
\end{equation}
where $\dot \lambda (t)=d\lambda(t) / dt$. This notation emphasizes that the work $W$ depends both on the specific trajectory followed by the system as well as on the particular protocol used to switch the control parameter. In path integral notation,  the Jarzynski equality can be written as:
\begin{equation}
\label{equ:Jarzynski_derivation}
e^{-\beta \Delta F} = \int {\cal D}x(t)\; {\cal P}[x(t), \lambda(t)] e^{-\beta W[x(t), \lambda(t)]}.
\end{equation}
Here, $\int {\cal D}x(t) ...$ denotes a summation over all trajectories and ${\cal P}[x(t),\lambda(t)]$ is the probability density of trajectory $x(t)$, along which the work $W[x(t), \lambda(t)]$ is performed during the switching process. The probability density ${\cal P}[x(t),\lambda(t)]$ includes the canonical distribution $\rho(x_0)=\exp[-\beta {\cal H}(x_0, \lambda_A)]/Q_{\lambda_A}$ of the initial conditions $x_0$. For deterministic dynamics, the trajectory $x(t)$ and the work $W[x(t), \lambda(t)]$ are fully described by the initial point in phase space $x_0$. In this case, the integral over all trajectories can be written as an integral over phase space rather than an integral over trajectory space. In the following, we will often omit the explicit dependence of $W[x(t)]$ and ${\cal P}[x(t)]$ on $\lambda (t)$ to simplify the notation. 

\subsection{Straightforward fast switching}

The Jarzynski equation (\ref{equ:Jarzynski_derivation}) justifies the following algorithm to estimate the free energy difference between the two equilibrium states $A$ and  $B$ corresponding to $\lambda_A$ and $\lambda_B$, respectively. First, one generates $N$ initial conditions canonically distributed with respect to ${\cal H}(x_0, \lambda_A)$. This can be done with Monte Carlo sampling or constant temperature molecular dynamics. From each of the $N$ initial condition $x_0^{(i)}$ one then generates a trajectory of length $\tau$ by integrating the appropriate equations of motion and calculates the work $W^{(i)}$ performed on the system along that trajectory. From this sample of $N$ trajectories an estimate  $\Delta \overline{F}_N$  of the free energy is then determined:
\begin{equation}
\Delta \overline{F}_N \equiv -k_{\rm B}T \ln  \frac{1}{N}\sum_{i=1}^{N} e^{-\beta W^{(i)}}.
\end{equation}
As the trajectory sample is finite, this free energy estimate contains errors \cite{BUSTAMANTE} that are discussed in detail in Sec. \ref{sec:error}. Due to the highly non-linear behavior of the exponential function these errors may be severe, often precluding an accurate free energy calculation particularly in the case of large switching rates \cite{jarTPS}. The reason for these large deviations is best perceived by considering the work distribution $P(W)$:
\begin{equation}
P(\tilde W) \equiv  \int\!\mathcal{D}x(t)\;{\cal P}[x(t)] \delta (\tilde W - W[x(t)]).
\end{equation}
In terms of $P(W)$, the Jarzynski equation can be rewritten as an integral over the work:
\begin{equation}
e^{-\beta \Delta F}=\int_{-\infty}^{\infty} dW P(W) e^{-\beta W}.
\label{equ:integral}
\end{equation}
In a fast switching simulation as described above, most trajectories have work values near the values for which $P(W)$ is a maximum. For large switching rates, $P(W)$  may be centered at large work values, such that for typical work values $\exp(-\beta W)$ is a very small number yielding a vanishing contribution to the integral in Eq. (\ref{equ:integral}). The work values leading to the important contributions to the integral, on the other hand, may occur very rarely \cite{JARZYNSKI_RAREEVENTS}. This situation, familiar from the calculation of chemical potentials with Widom's particle insertion method \cite{WIDOM} or from Zwanzig's perturbation approach \cite{Zwanzig}, may lead to extremely large statistical inaccuracies. These statistical difficulties can easily offset the advantage of using short, computationally inexpensive non-equilibrium trajectories.

\subsection{Work biased thermodynamic integration}

To avoid large statistical errors occurring in the straightforward application of the Jarzynski relation, Sun proposed a method based on a  thermodynamic integration in trajectory space \cite{SUN}. The basic idea of this method is to introduce a bias that favors the sampling of the rare but important trajectories that mostly contribute to the exponential work average.  This is done by introducing a free energy $\Delta \tilde F(\alpha)$ that depends on the parameter $\alpha$:  
\begin{equation}
e^{-\beta \Delta \tilde F(\alpha)} \equiv  \int\!\mathcal{D}x(t){\cal P}[x(t)]e^{-\beta \alpha W[x(t)]}.
\label{equ:Falpha}
\end{equation}
For $\alpha=0$, the parameter dependent free energy vanishes because ${\cal P}[x(t)]$ is normalized, $\Delta \tilde F(0)=0$. For $\alpha=1$, on the other hand, the parameter dependent free energy is identical to the original free energy,  $\Delta \tilde F(1)=\Delta F$. One can therefore calculate $\Delta F$ by taking the derivative of $\Delta \tilde F(\alpha)$ and integrating it from $0$ to $1$:
\begin{equation}
\label{equ:sunasintegral}
\Delta F = \int_0^1 d\alpha \frac{d \Delta \tilde F(\alpha)}{d\alpha}.
\label{equ:intalpha}
\end{equation}
To carry out this  procedure, which is the path space version of the thermodynamic integration method \cite{Kirkwood}, one needs the derivative of  $\Delta \tilde F(\alpha)$ with respect to $\alpha$:
\begin{equation}
\label{equ:sunderivative}
\frac{d \Delta \tilde F(\alpha)}{d\alpha}=\frac{\int\!\mathcal{D}x(t){\cal P}[x(t)] e^{-\beta \alpha W[x(t)]}W[x(t)]}{\int\!\mathcal{D}x(t){\cal P}[x(t)]e^{-\beta \alpha W[x(t)]}}.
\end{equation}
The right hand side of this equation can be viewed as the average $\langle W \rangle_\alpha$ of $W[x(t)]$ over the work biased path ensemble
\begin{equation}
{\cal P}_\alpha[x(t)]={\cal P}[x(t)]e^{-\beta \alpha
W[x(t)]} / Z_\alpha,
\label{equ:Pwork}
\end{equation}
where 
\begin{equation}
Z_\alpha=\int\!\mathcal{D}x(t){\cal P}[x(t)] e^{-\beta \alpha W[x(t)]}.
\end{equation}
In terms of the work average $\langle W \rangle_\alpha$, Eq.(\ref{equ:intalpha}) can be rewritten as:
\begin{equation}
\Delta F = \int_0^1 d\alpha \; \langle W \rangle_\alpha.
\label{equ:TI}
\end{equation}
To calculate the free energy difference $\Delta F$ using this equation it is first necessary to determine $\langle W \rangle_\alpha$ for different values of the parameter $\alpha$ between 0 and 1. Then, the integral on the right hand side is determined numerically from these values. As shown by Sun \cite{SUN}, path sampling techniques, originally developed to study rare but important events \cite{TPS1,TPS2}, can be used to sample the work biased ensemble and calculate the averages $\langle W \rangle_\alpha$.

To understand why the work biased thermodynamic integration yields accurate free energy estimates, it is instructive to consider the work distributions $P_\alpha(W)$ in the work biased ensembles:
\begin{eqnarray}
P_\alpha(\tilde W) &\equiv&  \int\!\mathcal{D}x(t)\;{\cal P}_\alpha[x(t)] \delta (\tilde W - W[x(t)]) \nonumber \\
 & = & \frac{P(\tilde W)\exp(-\beta \alpha \tilde W)}{\int dW
P(W)\exp(-\beta\alpha W)}
\label{equ:PW_work_biased}
\end{eqnarray}
For $\alpha=0$ the parameter dependent work distribution equals the original work distribution:
\begin{equation}
P_0(W)=P(W).
\end{equation}
In the other limit, at $\alpha=1$, the parameter dependent work distribution is 
proportional to the integrand in Eq. (\ref{equ:integral}): 
\begin{equation}
P_1(W)=P(W)e^{-\beta (W-\Delta F)}.
\end{equation}
For values of the parameter $\alpha$ between 0 and 1 the work distribution in the work biased ensemble is intermediate between these two limiting cases. Thus, both the typical and the dominant work values are generated with comparable likelihood as $\alpha$ is changed from 0 to 1 yielding a good statistical accuracy.

\subsection{Work biased umbrella sampling}

An alternative way to enhance the sampling of rare but important work values was recently proposed by Ytreberg and Zuckerman \cite{YTREBERG} and Ath\`enes \cite{ATHENES}. The basic idea of this approach, called single-ensemble nonequilibrium path-sampling by Ytreberg and Zuckerman, is to introduce an explicit {\em biasing function} (or {\em umbrella function}) $\pi[x(t)]$ depending on the path $x(t)$. With this umbrella function the Jarzynski equality (\ref{equ:jarzynski}) can be rewritten as:
\begin{eqnarray}
\label{equ:biasave}
e^{-\beta \Delta F} & = & \frac
{\int {\cal D}x(t)\; {\cal P}[x(t)]\pi[x(t)]\left[\frac{e^{-\beta W[x(t)]}}
{\pi[x(t)]}\right]}{\int {\cal D}x(t)\; {\cal P}[x(t)]\pi[x(t)] \left[\frac{1}{\pi[x(t)]}\right]} \nonumber \\
& = & \frac{\langle e^{-\beta W[x(t)]}/\pi[x(t)] \rangle_\pi}{\langle 1/\pi[x(t)]\rangle_\pi}. 
\end{eqnarray}
Here, $\langle \cdots \rangle_\pi$ denotes an average over the biased path ensemble
\begin{equation} 
{\cal P}_\pi [x(t)]=\frac{{\cal P}[x(t)]\pi[x(t)]}{\int {\cal D}x(t)\; {\cal P}[x(t)]\pi[x(t)]}. 
\end{equation}
For a path observable $A[x(t)]$ depending on the pathway $x(t)$ the average in the biased path ensemble is given by:
\begin{equation}
\langle  A[x(t)] \rangle_\pi\equiv \frac
{\int {\cal D}x(t)\; {\cal P}[x(t)]\pi[x(t)]A[x(t)] }
{\int {\cal D}x(t)\; {\cal P}[x(t)]\pi[x(t)]}.
\end{equation}
Such averages can be calculated using path sampling techniques \cite{YTREBERG,ATHENES,TPS1,TPS2}. Since the umbrella function is introduced to guide the sampling to the relevant work values, it suffices to let $\pi[x(t)]$ depend on the work only,  $\pi[x(t)]=\pi[W(x(t))]$. In this case, the biased average of a function $A[W(x(t))]$ that also depends on the work only can be written as
\begin{equation}
\langle  A \rangle_\pi\equiv \frac
{\int dW\; P(W)\pi(W)A(W)}
{\int dW\; P(W)\pi(W)}.
\end{equation}
This expression will be useful in the next section, in which we discuss the statistical errors occurring in fast switching simulations. 

In a work biased umbrella sampling simulation the choice of the umbrella function $\pi[x(t)]$ is crucial. To obtain a good accuracy in the free energy estimate, it is necessary that the bias function has a good overlap with the unbiased work distribution $P(W)$ as well as with the integrand of Eq. (\ref{equ:integral}), $P(W)\exp(-\beta W)$. Ytreberg and Zuckerman found that the umbrella function $\pi(W) = \exp(-\beta W/2)$ leads to considerable efficiency increases compared to the results of straightforward fast switching simulations. Another possibility would be to carry out umbrella sampling simulations with partially overlapping windows \cite{FRENKEL_SMIT}. 

Note that other non-Boltzmann sampling techniques such as flat histogram sampling \cite{WANG_LANDAU}, multicanonical sampling \cite{BERG}, or parallel tempering \cite{GEYER} may also be used to improve the convergence of the exponential average of the Jarzynski equation. These techniques, however, are not considered in the present article.

\section{Error Analysis}
\label{sec:error}

In this section we review the expressions required to estimate the statistical errors arising in fast switching simulations. While straightforward fast switching and work biased umbrella sampling can be treated in the same way, separate expressions are required for the work biased thermodynamic integration scheme.

\subsection{Straightforward fast switching and work biased umbrella sampling}

To set the notation we reproduce here the expressions derived in Ref. \cite{jarTPS}. All expressions are valid both for work biased umbrella sampling as well as for straightforward fast switching. The latter case follows by setting the umbrella function equal to unity, $\pi[x(t)] = 1$. 

For convenience we define 
\begin{equation}
X[x(t)]\equiv \frac{e^{-\beta W[x(t)]}}{\pi[x(t)]} \quad \mbox{and} \quad Y[x(t)]\equiv \frac{1}{\pi[x(t)]}.
\end{equation}
The free energy estimated according to Eq. (\ref{equ:biasave}) from $N$ pathways is  
\begin{equation}
\label{equ:dF_finite}
\Delta \overline{F}_N \equiv -k_{\rm B}T\ln \frac{\overline{X}_N}{\overline{Y}_N},
\end{equation} 
where $\overline{X}_N$ and $\overline{Y}_N$ are averages over the $N$ trajectories of the simulation,
\begin{equation}
\overline{X}_N\equiv\frac{1}{N}\sum_{i=1}^{N} X^{(i)} \quad \mbox{and} \quad  
\overline{Y}_N\equiv\frac{1}{N}\sum_{i=1}^{N} Y^{(i)}.
\end{equation}
Here, $X^{(i)}$ and $Y^{(i)}$ are the values of $X$ and $Y$ associated with the $i$-th path sampled from the biased ensemble, $P_\pi [x(t)]$.

Due to the non-linearity of the logarithm, this estimator for the free energy is biased, i.e., the average over many realizations $\Delta \overline{F}_N^{(j)}$ of the free energy estimator,
\begin{equation}
\langle \Delta \overline{F}_N \rangle\equiv \lim_{M\rightarrow \infty}\frac{1}{M}\sum_{j=1}^{M}\Delta \overline{F}_N^{(j)},
\end{equation}
differs from the true free energy difference. This deviation,   
\begin{equation}
\label{equ:bias}
b_N\equiv\langle \Delta \overline{F}_N \rangle-\Delta F,
\end{equation}
is called the bias. In addition to this systematic deviation, the estimator $\Delta \overline{F}_N$ is also affected by statistical errors which are quantified by the variance,
\begin{equation}
\sigma_N^2\equiv \langle [\Delta \overline{F}_N-\langle \Delta \overline{F}_N \rangle]^2\rangle=
\langle (\Delta \overline{F}_N)^2\rangle - \langle \Delta \overline{F}_N\rangle^2.
\end{equation}
The total mean squared deviation of the estimator $\Delta \overline{F}_N$ from the true free energy difference $\Delta F$ has contributions from the bias and the statistical error:
\begin{equation}
\epsilon_N^2 \equiv \langle [\Delta \overline{F}_N- \Delta F]^2\rangle=b_N^2+\sigma_N^2
\end{equation}

We now consider the limit of a large sample size $N$ and first calculate the bias $b_N$. For large $N$, the deviations $\delta \overline{X}_N$ and $\delta \overline{Y}_N$ from their averages $\overline{X}_N$ and $\overline{Y}_N$, 
\begin{equation}
\delta \overline{X}_N= \overline{X}_N - \langle X \rangle_\pi \quad \mbox{and} \quad 
\delta \overline{Y}_N= \overline{Y}_N - \langle Y \rangle_\pi,
\end{equation}
are small compared to $ \langle X \rangle_\pi$ and $ \langle Y \rangle_\pi$, respectively. Expansion of the logarithm in Eq. (\ref{equ:dF_finite}) into a Taylor series around $\langle X \rangle / \langle Y \rangle$ and truncation after the quadratic term (the linear terms vanish after averaging) yields the bias
\begin{equation} 
b_N = \frac{k_{\rm B}T}{2N}\left[
\frac{\langle (\delta X)^2 \rangle_\pi}{\langle X\rangle^2_\pi}-
\frac{\langle (\delta Y)^2 \rangle_\pi}{\langle Y\rangle^2_\pi}\right],
\label{equ:bias_approx}
\end{equation}
where the fluctuations $\delta X$ and $\delta Y$ are given by
\begin{equation}
\delta X \equiv X - \langle X \rangle_\pi \quad \mbox{and} \quad 
\delta Y \equiv Y - \langle Y \rangle_\pi.
\end{equation}
Interestingly, Eq. (\ref{equ:bias_approx}) implies that the bias can be made to vanish by choosing an appropriate umbrella function. 

For the variance $\sigma_N$ and the mean squared error $\epsilon_N$ the Taylor series  of the logarithm in Eq. (\ref{equ:dF_finite}) can be truncated after the linear term. In this approximation the bias vanishes and the variance and the mean squared error are identical:
\begin{eqnarray}
\epsilon_N^2 & = & \sigma_N^2= \frac{k_{\rm B}^2T^2}{N} \nonumber \\ 
& & \times \left[\frac{\langle (\delta X)^2 \rangle_\pi}{\langle X\rangle^2_\pi}+
\frac{\langle (\delta Y)^2 \rangle_\pi}{\langle Y\rangle^2_\pi}-2
\frac{\langle \delta X\delta Y \rangle_\pi}{\langle X\rangle \langle Y\rangle_\pi}\right].
\label{equ:epsilon}
\end{eqnarray}
In the absence of an umbrella function $\pi$ (or, in other words, for $\pi=1$) these expressions describe the errors encountered in straightforward fast switching simulations and are identical to the ones derived in Refs. \cite{BUSTAMANTE} and \cite{ZUCKERMAN_WOOLF}.

All terms occurring in the above expressions for the bias $b_N$, the variance $\sigma_N$, and the mean squared error $\epsilon_N$ can be calculated as integrals involving the work distribution $P(W)$:
\begin{eqnarray}
\label{equ:integral_f}
I &=& \int dW\; P(W)\pi(W),\\
\langle X \rangle_\pi & = & \frac{1}{I}\int dW\; P(W) \exp(-\beta W),\\
\langle Y \rangle_\pi & = & \frac{1}{I},\\
\langle X^2 \rangle_\pi & = & \frac{1}{I}\int dW\; P(W) \left[\frac{\exp(-2\beta W)}{\pi(W)}\right],\\
\langle Y^2 \rangle_\pi & = & \frac{1}{I}\int dW\; P(W)\left[\frac{1}{\pi(W)}\right],\\
\langle XY \rangle_\pi & = & \frac{1}{I}\int dW\; P(W) \left[\frac{\exp(-\beta W)}{\pi(W)}\right].
\label{equ:integral_l}
\end{eqnarray}
In section \ref{sec:results} we will use these equations to calculate the expected errors for two systems with analytically known work distributions.  

In the following we will also consider the errors in the free energy if the process is carried out in reverse direction, i.e., if the system starts from equilibrium initial conditions corresponding to the final value of the control parameter and the system evolves under a time-inverted switching protocol. In general, these errors differ from those of the forward process. However, in the case of umbrella sampling with bias  $\pi(W) = \exp(-\beta W /2)$ the errors of the forward and reverse process are identical. To show this we consider the work distribution $P_R(W)$ of the reverse process, which is related to the work distribution of the forward process by the Crooks identity \cite{CROOKS,gavin_jstatphys}
\begin{equation}
\label{equ:crooks}
P_R(-W) = P(W) e^{-\beta W+\beta \Delta F} .
\end{equation}
Insertion of Eq. (\ref{equ:crooks}) into Eqs. (\ref{equ:integral_f}) to (\ref{equ:integral_l}) yields
\begin{eqnarray}
\label{equ:expw2identity}
\frac{\langle X^2 \rangle_{{\pi},R}}{\langle X \rangle_{{\pi},R}^2} & = & \frac{\langle Y^2 \rangle_{{\pi},F}}{\langle Y \rangle_{{\pi},F}^2},\\
\frac{\langle Y^2 \rangle_{{\pi},R}}{\langle Y \rangle_{{\pi},R}^2} & = & \frac{\langle X^2 \rangle_{{\pi},F}}{\langle X \rangle_{{\pi},F}^2},\\
\frac{\langle XY \rangle_{{\pi},R}}{\langle X \rangle_{{\pi},R} \langle Y \rangle_{\pi,R}} & = & \frac{\langle XY \rangle_{{\pi},F}}{\langle X \rangle_{{\pi},F} \langle Y \rangle_{{\pi},F}}.
\end{eqnarray}
Here, $\langle .. \rangle_{{\pi},F}$ and $\langle .. \rangle_{{\pi},R}$ denote averages with  umbrella function ${\pi}(W)$ for the forward and the reverse process, respectively. Thus, the fluctuations of $X$ in the forward process are identical to the fluctuations of $Y$ in the reverse process and vice versa. The correlation term is the same for both directions. Inserting these results into Eq. (\ref{equ:epsilon}) one finally obtains  
\begin{equation}
\label{equ:reverseforward}
\epsilon^2_{N \textit{,} F} = \epsilon^2_{N \textit{,} R},
\end{equation}
i.e., the mean squared errors for the forward and reverse process are identical. In the same way one can show that also for the $1/P$-bias the errors are the same in forward and reverse direction.

For straightforward fast switching Eq. (\ref{equ:epsilon}) is very similar to an expression obtained recently by Jarzynski that relates the number of trajectories required to obtain acceptable accuracy to the dissipative work of the reverse process \cite{JARZYNSKI_RAREEVENTS}. To establish this similarity we use Eq. (\ref{equ:epsilon}) to determine the number $N_{kT}$ of trajectories required to obtain a statistical error of thermal magnitude $k_{\rm B}T$:
\begin{equation}
N_{kT}=\frac{\int dW P(W) \exp(-2\beta W)}{\exp(-2\beta \Delta F)}-1.
\label{equ:number}
\end{equation}
Using the Crooks identity, Eq. (\ref{equ:number}) can be written as
\begin{equation}
N_{kT}=\frac{\int dW P_R(-W) \exp(-\beta W)}{\exp(-\Delta F)},
\end{equation}
where we have neglected the term -1 on the right hand side of Eq. (\ref{equ:number}) (in any interesting case $N$ is much larger than 1). Changing the integration variable from $W$ to $-W$ one obtains:
\begin{eqnarray}
N_{kT}&=&\int dW P_R(W) \exp[\beta (W+\Delta F)] \nonumber \\
&=&\langle \exp(\beta W_R^d) \rangle_R,
\label{equ:diss}
\end{eqnarray}
where $W_R^d=W+\Delta F$ is the dissipative work for the reverse transformation and $\langle \cdots \rangle_R$ indicates an average over realizations of the reverse process. This equation implies that the number of fast switching trajectories required for an accurate free energy estimation depends on the work dissipated during the reverse process, as noted earlier by Jarzynski \cite{JARZYNSKI_RAREEVENTS}. Remarkably, for highest accuracy of the free energy estimate the process should be carried out in the direction that yields the larger exponential average of the dissipative work \cite{JARZYNSKI_RAREEVENTS}.  Of course, the same conclusion can be drawn from 
\begin{equation}
N_{kT}=\langle\exp(-2\beta W^d)\rangle,
\end{equation}
which also simply follows from Eq. (\ref{equ:number}) and where $W^d=W-\Delta F$ is the dissipative work for the forward process.

To simplify Eq. (\ref{equ:diss}), one is tempted to replace the average of the exponential with the exponential of the average \cite{JARZYNSKI_RAREEVENTS},
\begin{equation}
\langle \exp(\beta W_R^d) \rangle_R \approx \exp(\beta \overline W_R^d), 
\label{equ:Wdiss}
\end{equation}
where $ \overline W_R^d=\int  dW P_R(W) (W+\Delta F)$ is the average dissipative work for the reverse process. This approximation, however, is not valid in general. As shown for an illustrative example in Sec. \ref{sec:pulled}, it is likely to underestimate the required number of trajectories that may be off by orders of magnitude from the true value. 

\subsection{Work biased thermodynamic integration}

According to Eq. (\ref{equ:TI}), Sun's work biased thermodynamic integration method requires the calculation of an integral over the parameter $\alpha$. To determine this integral, the average $\langle W \rangle_\alpha$ is first calculated for $L$ discrete values $\alpha_j$ of the parameter $\alpha$ between 0 and 1 in $L$ separate simulations. Then, the integration is carried out numerically using for instance the trapezoidal or Simpson's rule. Imagine now that the average work $\langle W \rangle_\alpha$ is approximated by averaging over $M$ trajectories at each value of the parameter $\alpha$:
\begin{equation}
\langle W \rangle_\alpha\approx  \overline{W}_M(\alpha)=\frac{1}{M} \sum_{i=0}^M W^{(i)}.
\end{equation}
Here, the argument $\alpha$ in $\overline{W}_M(\alpha)$ indicates that the trajectories over which the average is carried out are sampled from the trajectory ensemble corresponding to the parameter $\alpha$.
Using the trapezoidal integration rule the estimate for the free energy is then given by
\begin{equation}
\Delta \overline{F}_N=\frac{1}{L}\sum_{j=1}^{L}\overline{W}_M(\alpha_j),
\end{equation} 
where $N=M L$ is the total number of trajectories generated in the simulation. Neglecting errors in the numerical integration over $\alpha$ this free energy estimate is bias free:
\begin{equation}
\langle \Delta \overline{F}_N\rangle =
\frac{1}{L}\sum_{j=1}^{L}\langle \overline{W}_M(\alpha_j)\rangle=
\frac{1}{L}\sum_{j=1}^{L}\langle W\rangle_{\alpha_j}
= \Delta F,
\end{equation} 
where we have used that $\langle \overline{W}_M(\alpha_j)\rangle=\langle W\rangle_{\alpha_j}$. In the above equation the angular brackets $\langle \cdots \rangle$ denote an average over many independent realizations of the thermodynamic integration procedure. 

To determine the variance $\sigma_N^2$ of the free energy estimate, which in this case is equal to the mean squared error $\epsilon_N^2$, we note that the variance of a sum of random variables equals the sum of the variances of the individual variables. Accordingly, the variance of $\Delta \overline{F}_N$ is given by 
\begin{equation}
\label{equ:variance_dF}
\sigma_N^2 = \frac{1}{L^2}\sum_{j=1}^{L} \sigma^2_{\overline{W}}(\alpha_j),
\end{equation}
where
\begin{equation}
\sigma^2_{\overline{W}}(\alpha_j)=\langle [\overline{W}_M(\alpha_j)-\langle W\rangle_{\alpha_j}]^2\rangle
\end{equation}
is the variance of the average $\overline{W}_M(\alpha_j)$ at parameter value $\alpha_j$. For statistically independent work values this variance is given by
\begin{equation}
\sigma^2_{\overline{W}}(\alpha_j)=\frac{1}{M}[\langle W^2\rangle_{\alpha_j}-\langle W\rangle^2_{\alpha_j}]
\end{equation}
Rewriting the sum over $j$ in Eq. (\ref{equ:variance_dF}) as an integral over $\alpha$ we finally obtain:
\begin{equation}
\label{equ:var_dFN}
\sigma^2_N
  =  \frac{1}{N}\int_0^1 \left[\langle W^2\rangle_{\alpha}-\langle W\rangle^2_{\alpha}\right] d\alpha.
\end{equation}
Note that here we have assumed that at each value of the parameter $\alpha$, the same number $M$ of trajectories are generated. To minimize the variance $\sigma_N^2$, however, it is advantageous to sample more trajectories at values of $\alpha$ where the work variance is large and less where it is small. For simplicity we do not consider this case here.

For a given work distribution $P(W)$ the work variance in the work biased ensemble can be calculated from 
\begin{equation}
\langle W \rangle_\alpha = \frac{\int dW P(W) e^{-\alpha \beta W}W}{\int dW P(W) e^{-\alpha \beta W}}
\end{equation}
and 
\begin{equation}
\langle W^2 \rangle_\alpha = \frac{\int dW P(W) e^{-\alpha \beta W}W^2}{\int dW P(W) e^{-\alpha \beta W}}.
\end{equation}
By numerical integration of Eq. (\ref{equ:var_dFN}) one can then calculate the variance $\sigma_N^2$ of the free energy estimate $\Delta \overline{F}_N$.

Interestingly, also the error of the thermodynamic integration method for the reverse process is identical to  that for the forward process. The average of an arbitrary function $A(W)$ for a particular $\alpha$ of the reverse process is
\begin{equation}
\label{equ:reverseaverage}
\langle A(W) \rangle_{\alpha \textit{,} R} = \frac{\int dW P_R(W) e^{-\alpha \beta W}A(W)}{\int dW P_R(W) e^{-\alpha \beta W}}.
\end{equation}
Inserting the Crooks identity, Eq. (\ref{equ:crooks}), into Eq. (\ref{equ:reverseaverage}) we obtain 
\begin{eqnarray}
\langle A(W) \rangle_{\alpha \textit{,} R} &=& \frac{\int dW P(W) e^{-\beta (1-\alpha) W}A(-W)}{\int dW P(W) e^{-\beta (1-\alpha)W}}  \\ &=& \langle A(-W) \rangle_{1-\alpha \textit{,} F}.
\end{eqnarray}
Thus, the work variance for the forward process at $\alpha$ is equal to the work variance for the reverse process at $(1-\alpha)$:
\begin{equation}
\label{equ:workaveragesreverse}
\langle W^2 \rangle_{\alpha \textit{,} R}- \langle W \rangle^2_{\alpha \textit{,} R} = 
\langle W^2 \rangle_{1-\alpha \textit{,} F} - \langle W \rangle^2_{1-\alpha \textit{,} F}
\end{equation}
Integration over $\alpha$ according to Eq. (\ref{equ:var_dFN}) with a change of variables from $\alpha$ to $1-\alpha$ in the reverse case then demonstrates that the statistical errors for the forward and reverse process are indeed equal:
\begin{equation}
\sigma^2_{N \textit{,} F} = \sigma^2_{N \textit{,} R}.
\end{equation}

\section{Efficiency}
\label{sec:efficiency}

While the error obtained from a simulation with $N$ trajectories is an interesting quantity, the criterion most relevant for the practitioner is the computational cost required to achieve a given accuracy in the free energy. For the estimation of this cost one needs to take into account that long trajectories are computationally more expensive than short ones. In order to do that we note that for all algorithms studied here the error in the free energy can be written as:
\begin{equation}
\epsilon^2_N=\frac{\kappa^2}{N},
\end{equation}
where $\kappa$ is a constant that does not depend on the number of trajectories used in the calculation. Hence the number $N_{kT}$ of trajectories required to obtained an accuracy of $k_{\rm B}T$ in the free energy is given by 
\begin{equation}
N_{kT} = \frac{\kappa^2}{k^2_{\rm B}T^2}.
\end{equation}
Here, we have set the target error to a value of $k_{\rm B}T$ because variations of the free energy much smaller than $k_{\rm B}T$ are irrelevant in most cases.

Since the computational cost of a trajectory is proportional to its length $\tau$ a meaningful definition of the computational cost is given by
\begin{equation}
C_{CPU}= N_{kT} \tau= \frac{\kappa^2}{k^2_{\rm B}T^2} \tau,
\label{equ:CCPU}
\end{equation}
$C_{CPU}$ is the computational cost of a free energy calculation with accuracy $k_{\rm B}T$ measured in units of the computational cost to generate a trajectory of length $\tau=1$. While for straightforward fast switching and umbrella sampling the computational cost is
\begin{equation}
C_{CPU}=\tau \left[\frac{\langle (\delta X)^2 \rangle_\pi}{\langle X\rangle^2_\pi}+
\frac{\langle (\delta Y)^2 \rangle_\pi}{\langle Y\rangle^2_\pi}-2
\frac{\langle \delta X\delta Y \rangle_\pi}{\langle X\rangle_\pi \langle Y\rangle_\pi}\right],
\end{equation}
for the work biased thermodynamic integration scheme is given by:
\begin{equation}
C_{CPU}=\beta^2 \tau \int_0^1 (\langle W^2\rangle_{\alpha}-\langle W\rangle^2_{\alpha})d\alpha.
\end{equation}

\section{Results}
\label{sec:results}

If the work distribution $P(W)$ is known for a specific switching process, the equations of the previous  sections can be used to determine the expected accuracy of the free energy calculated with the methods of Sec. \ref{sec:methods}. Here we do that for two models, for which the work distributions are known analytically. In particular, we study the computational efficiency of these methods in terms of the computing time required to obtain a given accuracy in the calculated free energy.

\subsection{Particle pulled through a viscous fluid}
\label{sec:pulled}

\subsubsection{Model and work distribution}

\begin{figure}[h]
\centerline{\includegraphics[width=7.0cm]{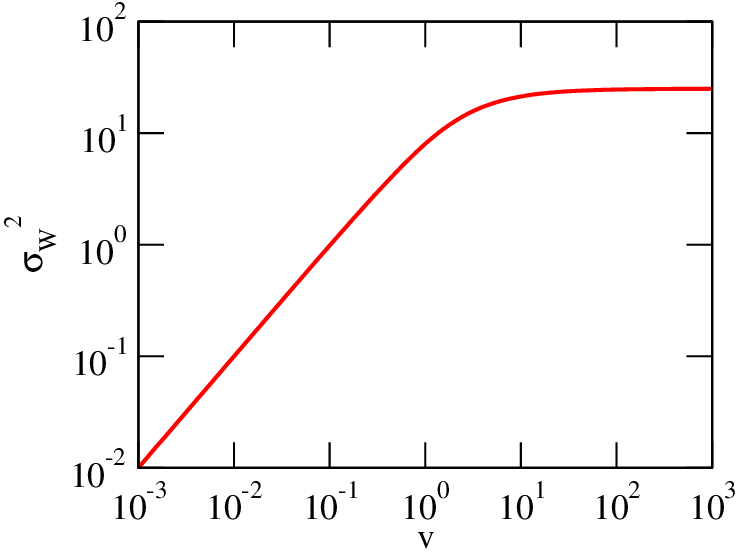}}
\caption{Work variance $\sigma_W^2$ for a particle pulled through a viscous fluid by a harmonic trap moving at constant speed $v$. The average work is given by $\overline W  = \beta \sigma^2_W / 2$.}
\label{fig:sig2}
\end{figure}

The first model we consider is a particle in one dimension pulled through a viscous liquid  at temperature $T$  by a harmonic trap with force constant $k$ moving at constant speed $v$ for a certain distance $L$. The translation of the trap by the distance $L$ takes a time $\tau=L/v$. The motion of the particle is modelled by a Langevin equation in the overdamped limit with friction constant $\gamma$:
\begin{equation}
\frac{dx}{dt}= -\frac{k}{\gamma}(x-vt) + \eta,
\end{equation} 
where $x$ is the position of the particle and $\eta$ is a Gaussian random variable uncorrelated in time, $\langle \eta (t) \eta (t') \rangle = (2k_{\rm B}T / \gamma) \delta (t-t')$. For this model, Mazonka and Jarzynski have computed the work distribution analytically finding a Gaussian distribution \cite{MAZONKA_JARZYNSKI}: 
\begin{equation}
P(W)=\frac{1}{\sqrt{2\pi} \sigma_W} \exp\left[-\frac{(W-\overline{W})^2}{2\sigma_W^2}\right]
\end{equation}
with mean
\begin{equation}
\overline{W} = \gamma L v \left[1+\frac{\gamma v}{k L}\left(e^{-kL /v \gamma}-1 \right) \right] 
\end{equation}
and variance
\begin{equation}
\sigma_W^2 = 2 \frac{\gamma L v}{\beta}  \left[1+\frac{\gamma v}{kL}\left(e^{-kL /v \gamma}-1 \right) \right]=\frac{2}{\beta}\overline{W}.
\label{equ:Wave}
\end{equation}
Since the free energy of the system does not depend on the position of the harmonic trap, $\Delta F = 0$ for this transformation. The average work, however, is larger than zero due to the energy dissipated into the bath and only in the limit of infinitely slow switching does the average work vanish. For slow switching, i.e., for $v \ll kL / \gamma$ or $\tau \gg \gamma / k$, the average work is proportional to the pulling velocity, $\overline{W}=\gamma L v$, as predicted by linear response theory. In the limit of infinitely fast switching, i.e. for $v \gg k L / \gamma$ or $\tau \ll \gamma /k$, the average work converges to a constant value, $\lim_{v \rightarrow \infty} \overline {W}=kL^2/2$. Accordingly, the variance $\sigma_W^2=2k_{\rm B}T\gamma L v$ for slow switching. In the fast switching regime the variance reaches a constant value, $\sigma_W^2=k _{\rm B}TkL^2$ (see Fig. \ref{fig:sig2}). All results shown in this section were obtained for $k_{\rm B}T=1$, $\gamma=1$, $k=1$ and $L=5$.

\subsubsection{Efficiency}

For this model we have calculated the expected errors for straightforward fast switching (i.e., $\pi_1 (W)=1$) as well as for the two bias functions $\pi_2(W)=\exp(-\beta W / 2)$, and $\pi_3(W)=1/P(W)$. The bias function $\pi_2(W)=\exp(-\beta W / 2)$ was suggested by Ytreberg and Zuckerman and is designed to improve the sampling of low work values \cite{YTREBERG}. Some attention is needed when considering the umbrella function $\pi_3(W)$. Since this function biases the simulation in a way to make the work distribution flat, $\pi_3(W)$ can have the form $1/P(W)$ only in a range with finite size. We therefore define 
\begin{equation}
\pi_3(W)=\left\{
\begin{array}{lcl}
1 / P(W_{\rm min}) & \text{for} & W \le W_{\rm min}, \\
1/P(W) &  \text{for} & W_{\rm min} < W < W_{\rm max}, \\
1 / P(W_{\rm max}) & \text{for} & W \ge W_{\rm max}. 
\end{array}
\right.
\end{equation}
The limits for this range, $W_{\rm min}$ and $W_{\rm max}$, need to be selected such that all important work values are included in the sampling. Of course, in the general case the work distribution $P(W)$ required to implement the umbrella function $\pi_3(W)=1/P(W)$ is not known in advance. However, methods such as flat histogram sampling \cite{WANG_LANDAU} can be used to determine this bias iteratively. Here, we study this case in order to evaluate the efficiency of the work biased umbrella sampling approach if a good umbrella function is available.

\begin{figure}[h]
\centerline{\includegraphics[clip=true,width=7.0cm]{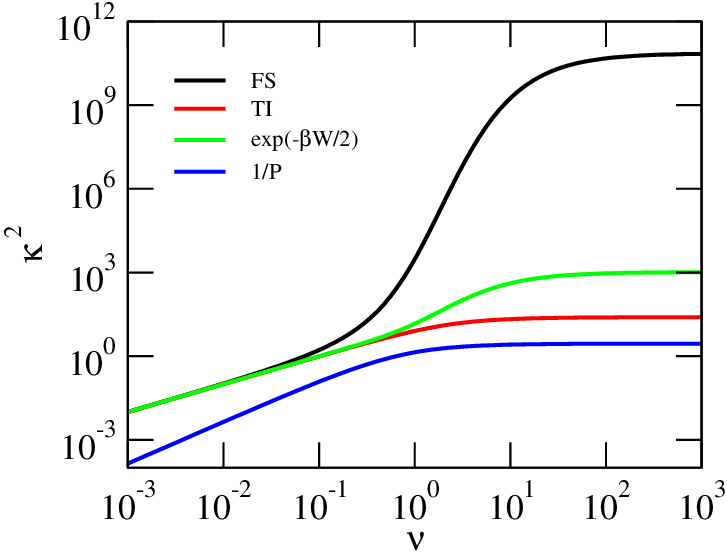}}
\caption{Error measure $\kappa^2=\epsilon^2_N N$ for the particle pulled through a viscous fluid as a function of switching rate $v$ obtained for straightforward fast switching (FS), work biased thermodynamic integration (TI), and work biased umbrella sampling with the umbrella functions $\pi=\exp(-\beta W / 2)$ and $\pi=1/P(W)$. Here, $k_{\rm B}T=1$, $\gamma=1$, $k=1$ and $L=5$. }
\label{fig:gauss_err}
\end{figure}

For the Gaussian work distribution of the pulled harmonic oscillator and the bias functions considered here all integrals in Eqs. (\ref{equ:integral_f}) to (\ref{equ:integral_l}) can be computed analytically. These analytical expressions can then be used to calculate the expected errors for straightforward fast switching,
\begin{equation}
\label{equ:eps_jar}
\epsilon^2_N=\frac{k^2_{\rm B}T^2}{N}\left(e^{\beta^2 \sigma_W^2}-1 \right),
\end{equation}
work biased umbrella sampling with the exponential umbrella function $\pi(W)=\exp(-\beta W/2)$,
\begin{equation}
\label{equ:eps_pi1}
\epsilon^2_N=\frac{k^2_{\rm B}T^2}{N}2e^{\beta^2 \sigma_W^2/4}
\left(1-e^{-\beta^2 \sigma_W^2/2}\right),
\end{equation}  
and with the umbrella function $\pi_3(W)=1/P(W)$,
\begin{equation}
\label{equ:eps_pi2}
\epsilon^2_N=\frac{k^2_{\rm B}T^2}{N}\frac{U}{\sqrt{\pi}\sigma_W}
\left(1-e^{-\beta^2 \sigma_W^2/4}\right).
\end{equation}  
Here, the interval size $U=W_{\rm max} - W_{\rm min}$ depends on the selection of the limits $W_{\rm max}$ and $W_{\rm min}$. As a measure for the interval $U$ in which the work distribution is required to be flat we use the distance between the maximum of the work distribution $P(W)$ and the maximum of the Jarzynski integrand $P(W)\exp(-\beta W)$. For the Gaussian work distributions considered here this distance is $\beta \sigma_W^2$ and we set $U =\beta \sigma_W^2$ accordingly. With this choice, the statistical error of the $1/P$ bias becomes: 
\begin{equation}
\label{equ:eps_pi2U}
\epsilon^2_N=\frac{k^2_{\rm B}T^2}{N}\frac{\beta \sigma_W}{\sqrt{\pi}}
\left(1-e^{-\beta^2 \sigma_W^2/4}\right),
\end{equation}  
The expected errors are shown in Fig. \ref{fig:gauss_err} as a function of the switching rate $v$. Since in all expressions for the error the number $N$ of trajectories factors out in a simple way, we have represented $\kappa^2=\epsilon^2_N N$ as a function of $v$ in this figure. The number $N_{kT}$ of trajectories required to obtaine an error of $k_{\rm B}T$ is related to $\kappa^2$ simply by $N_{kT} = \kappa^2 / k_{\rm B}T$.

We also calculated the expected errors for Sun's work biased thermodynamic integration scheme. Direct evaluation of the work distribution $P_\alpha(W)$ [see Eq. (\ref{equ:PW_work_biased})] in the work biased ensemble shows that for a Gaussian distribution $P(W)$ the variance of the work biased distribution $P_{\alpha}(W)$ is the same for all values of $\alpha$. In this case Eq. (\ref{equ:var_dFN}) yields 
\begin{equation}
\label{equ:eps_ti}
\epsilon^2_N=\frac{1}{N} (\langle W^2\rangle-\langle W\rangle^2)=\frac{1}{N} \sigma_W^2.
\end{equation}
While the variance in the work biased ensemble does not change as a function of $\alpha$, the average work $\langle W \rangle_\alpha$ does:
\begin{equation}
\langle W \rangle_\alpha=\langle W \rangle-\alpha\beta \sigma_W^2
\end{equation}
The expected errors for the work biased thermodynamic integration method are also shown in Fig. \ref{fig:gauss_err} as a function of the switching rate $v$. While for straightforward fast switching and for umbrella sampling with exponential bias the mean squared error grows exponentially with the dissipative work,  it increases with $\sqrt{\overline{W}}$ for the $1/P$ bias and linearly for thermodynamic integration.

As can be seen in Fig.  \ref{fig:gauss_err}, in the slow switching regime the errors decrease with decreasing $v$  in all four cases. To linear approximation in $v$, the error for straightforward fast switching, work biased umbrella sampling with the exponential bias $\pi(W)=\exp(-\beta W/2)$, and work biased thermodynamic integration is identical and given by 
\begin{equation}
\epsilon^2_N=\frac{\sigma_W^2}{N}=\frac{2\gamma k_{\rm B}TLv}{N}.
\end{equation}   
For work biased umbrella sampling with bias $\pi(W)=1/P(W)$ the error is smaller and the dependence on $v$ stronger:
\begin{equation}
\epsilon^2_N=\frac{\beta\sigma_W^3}{4\sqrt{\pi}N}=\frac{(2\gamma Lv)^{3/2} \sqrt{k_{\rm B}T}}{4\sqrt{\pi}N}.
\end{equation}  
As one proceeds to larger switching rates, the error grows most rapidly for straightforward fast switching. For work biased umbrella sampling with the exponential umbrella function the growth is less pronounced and occurs at higher switching rates. For the two remaining cases (work biased umbrella sampling with $\pi=1/P(W)$ and work biased thermodynamic integration) the growth of the error slows down rather then speeding up with increasing switching rate. Since the work variance becomes constant for large $v$, the error reaches a constant value in the fast switching regime in all four cases. 
In this limit we obtain
\begin{equation}
\label{equ:eps_jar_pull}
\epsilon^2_N=\frac{k^2_{\rm B}T^2}{N}e^{\beta^2 L^2 k}
\end{equation}
for straightforward fast switching,
\begin{equation}
\label{equ:eps_pi1_pull}
\epsilon^2_N=\frac{L^2k}{N}
\end{equation}  
for work biased thermodynamic integration,
\begin{equation}
\label{equ:eps_pi2_pull}
\epsilon^2_N=\frac{k^2_{\rm B}T^2}{N}2e^{\beta^2L^2k/4}
\end{equation}  
for work biased umbrella sampling with the exponential umbrella function $\pi(W)=\exp(-\beta W/2)$ and
\begin{equation}
\label{equ:eps_pi3_pull}
\epsilon^2_N=\frac{k^2_{\rm B}T^2}{N}\frac{\sqrt{\beta L^2k}}{\sqrt{\pi}},
\end{equation}  
in the case where the umbrella function $\pi(W)=1/P(W)$. Note, in Fig. \ref{fig:gauss_err}, that these constant values may differ by orders of magnitude.

\begin{figure}[h]
\centerline{\includegraphics[clip=true,width=7.0cm]{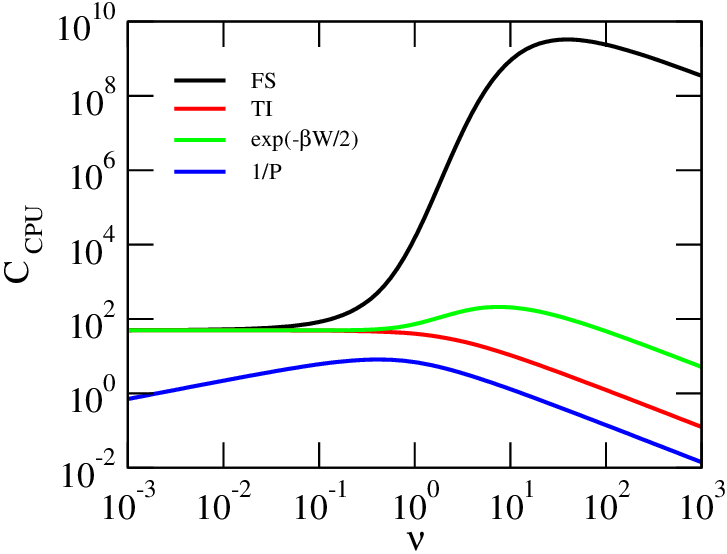}}
\caption{Computational cost $C_{\rm CPU}=\kappa^2\tau$ for the particle pulled through a viscous fluid as a function of switching rate $v$ obtained for straightforward fast switching (FS), work biased thermodynamic integration (TI), and work biased umbrella sampling with the umbrella functions $\pi=\exp(-\beta W / 2)$ and $\pi=1/P(W)$. Here, $k_{\rm B}T=1$, $\gamma=1$, $k=1$ and $L=5$.}
\label{fig:gauss_err_nu}
\end{figure}

The computational cost $C_{CPU}$ determined for the four algorithms studied here is shown as a function of the switching rate in Fig. \ref{fig:gauss_err_nu}. In the slow switching regime the computational cost becomes constant for straightforward fast switching, work biased umbrella sampling with exponential bias and work biased thermodynamic integration. This behavior indicates that in these cases running few long trajectories or many shorter trajectories yields the same efficiency. The situation is different for work biased umbrella sampling with the $1/P(W)$ umbrella function. In this case, $C_{\rm CPU}$ keeps decreasing for decreasing $v$ even in the slow switching regime. For larger switching rates straightforward fast switching and work biased umbrella sampling with exponential bias become very inefficient. In the other two cases, increasing the switching rate enhances the efficiency. With the appropriate bias the optimum efficiency is obtained in the instantaneous switching limit, in which case the path based sampling methods reduce to ordinary umbrella sampling and thermodynamic integration. Thus, conventional free energy calculation methods are superior to fast switching methods for the model studied in this section. Similar results have been obtained earlier for more complex models \cite{jarTPS,SUN}.
 
In addition to the expected error also the bias $b_N$ is an important quantity determining the accuracy of a free energy calculation. According to Eq. (\ref{equ:bias_approx}) the bias can be calculated using the expressions in Eqs. (\ref{equ:integral_f}) to (\ref{equ:integral_l}). For the pulled particle with Gaussian work distribution one obtains 
\begin{equation}
b_N=\frac{k_{\rm B}T}{2N}(e^{\beta^2\sigma_W^2}-1).
\end{equation}
in the case of straightforward fast switching. As can be seen by direct evaluation of Eq.  (\ref{equ:bias_approx}), all other algorithms are bias free, i.e., $b_N=0$, if the work distribution is  Gaussian.    

It is interesting that in the case of work biased sampling with exponential umbrella function, the property of vanishing bias is only given for the particular form $\pi(W)=\exp(-\beta W/2)$. To be more specific, one can consider umbrella functions $\pi(W)=\exp(-\lambda \beta W)$, where $\lambda$ is a real number between 0 and 1. In this case, error and bias are given by
\begin{eqnarray}
\epsilon^2_N & = &\frac{k_{\rm B}^2T^2}{N} \nonumber \\
& & \hspace{-0.5cm}\times \left[
e^{(1-\lambda)^2\beta^2\sigma_W^2}+e^{\lambda^2\beta^2\sigma_W^2}-2e^{-\lambda(1-\lambda)\beta^2\sigma_W^2}\right]
\end{eqnarray}
and
\begin{equation}
b_N=\frac{k_{\rm B}T}{2N}\left[
e^{(1-\lambda)^2\beta^2\sigma_W^2}-e^{\lambda^2\beta^2\sigma_W^2}\right],
\end{equation}
respectively. The error $\epsilon_N$ from the above equation is a minimum for $\lambda=1/2$. For this value of the parameter $\lambda$ the bias vanishes. In this sense, the umbrella function $\pi(W)=\exp(-\beta W/2)$ is the optimum umbrella function among all exponential functions of the form $\exp(-\lambda \beta W)$. Note, however, that this property holds strictly only for processes with Gaussian work distribution.  

We now briefly return to the question of whether in expression (\ref{equ:diss}) for the number of trajectories required to obtain an error of about $k_{\rm B}T$ the average of the exponential can be safely replaced by the exponential of the average (see Eq. \ref{equ:Wdiss}). For the process considered in this section, the dissipative work as well as the work distributions are the same in the forward and reverse direction. Furthermore, $W^d=W$ because $\Delta F=0$. Hence, Eq. (\ref{equ:diss}) reduces to:
\begin{equation}
N= \exp(\beta^2 \sigma_W^2)=\exp(2\beta\overline W),
\end{equation} 
where we have used Eq. (\ref{equ:Wave}) which is valid only for Gaussian work distributions. Thus, in this case, the estimate $N\approx \exp(\beta \overline W)$ may be off by orders of magnitude. If, for instance, the correct estimate is $N=10.000$, replacing the average of the exponential with the exponential of the average yields an estimate too low by a factor of $100$.

\subsection{Expansion of ideal gas}
\label{sec:Lua_Grosberg}

\subsubsection{Model and work distribution}

In this section, we will determine the efficiency of fast switching methods for the expansion of an ideal gas in a cylinder with moving piston. The work distribution for this process has been determined analytically by Lua and Grosberg \cite{GROSBERG_LUA_PEDAGOGICAL, LUA_Illustration}, who demonstrated that the Jarzynski identity holds exactly for arbitrary piston speed. In the limit of very rapid expansion this seems surprising as almost no gas particles collide with the piston performing work on it. Nevertheless, as shown by Lua and Grosberg, the tails of the Maxwell-Boltzmann distribution provide a sufficient number of fast moving particles that perform exactly the amount of work required for the Jarzynski identity to be valid.

The reason why we study the ideal gas expansion here is that its work distribution is non-Gaussian. As the system is not coupled to a heat bath during the transition, the work distribution remains non-Gaussian even in the limit of a  piston that moves infinitely slowly. In this case, the distribution is a rapidly fluctuating function with an exponential envelope. In the fast switching limit, the work distribution acquires a growing delta-peak at zero work and small contributions at large work values. These large but rare work values are the dominant contributions to the exponential work average.

In their work, Lua and Grosberg \cite{GROSBERG_LUA_PEDAGOGICAL, LUA_Illustration} considered a single ideal gas particle in a cylinder with cross section area $A$ sealed by a piston on one side. The volume available to the gas particle is $V=AL_0$, where $L_0$ is the initial length of the cylinder. Starting from initial conditions distributed canonically with temperature $T=(k_B \beta)^{-1}$, the volume of the cylinder is then increased by moving the piston from position $L_0$ to a new position $L$ in a time $\tau$ with constant velocity $v_p=(L-L_0)/\tau$. The ideal gas particle is assumed to collide elastically with the walls of the cylinder and the moving piston. At each elastic collision of the particle with the piston, a certain amount of kinetic energy is transferred from the particle to the piston. This adiabatic case is in contrast to that considered in Ref. \cite{BAULE}, where the system is coupled to a thermal bath and the velocities are taken from an equilibrium distribution at each collision with the piston. The sum of these energy transfers during one expansion is the work $W$ carried out by the gas. Note that here, as in Lua and Grosberg's work, $W$ denotes the work done {\it by} the system and not {\it on} the system. Accordingly, we let $\Delta F=F(A)-F(B)$ denote the free energy difference between the initial and the final state. The maximum work theorem then implies that the average work performed by the system is bounded from above by the free energy difference, $\langle W \rangle \le \Delta F$. With this sign convention, the Jarzynski equality is given by $\langle \exp(\beta W)\rangle = \exp(\beta \Delta F)$.

Lua and Grosberg have analyzed the statistics of the particle-piston collisions in detail for the special case of inverse temperature $\beta = 1$, mass of the particle $m = 1$ and switching time $\tau = 1$. In this paper we keep the explicit dependence on all variables and obtain the following work distribution :
\begin{eqnarray}
\label{equ:PW_gas}
\hspace{-0.5cm}P(W,L_0,v_p,\beta) &=&P_0\delta (W) +\nonumber \\
&& \hspace{-2.5cm}\sqrt{\frac{m\beta}{2 \pi}} \frac{1}{m n v_p} e^{-\frac{m \beta}{2} (nv_p + \frac{W}{2mnv_p})^2} f(W,L_0,v_p,\beta).
\end{eqnarray}
where
\begin{equation}
P_0 = \frac{1}{L_0} \sqrt{\frac{m\beta}{2 \pi}}  \int_0^{L_0} dx \int_{-(\frac{L_0}{\Delta L} + 1)v_p}^{(\frac{L_0}{\Delta L} + 1)v_p} dv e^{-\frac{m \beta}{2}(v-\frac{x v_p}{\Delta L})^2}
\end{equation}
is the probability that the particle does not collide with the moving piston during the expansion process. This probability increases with increasing piston velocity $v_p$ leading to a growing delta-peak at $W=0$ in the work distribution. As shown by Lua and Grosberg, the number $n$ of collisions between the moving ideal gas particle and the piston is completely determined by the work $W$ done by the gas during the entire expansion and is given by:
\begin{equation}
\label{equ:numberofbounces}
n(W,L_0,v_p) = \left\lfloor \frac{1}{2}+\frac{1}{2}\sqrt{1+\frac{2W}{mv_p^2(\frac{2L_0}{\Delta L}+1)}}\right\rfloor,
\end{equation}
where $\lfloor \cdots \rfloor$ indicates the floor function that gives the largest integer less or equal to the argument. The function $f(W,L_0,v_p,\beta)$ appearing in Eq. (\ref{equ:PW_gas}), called the overlap factor by Lua and Grosberg, is given by  the piecewise linear function
\begin{widetext}
\begin{equation}
\mbox{$f(W)=$} \left\{ \begin{array}{llrllll}
-(n-1)(1 + \frac{\Delta L}{2L_0}) + \frac{W}{4mnv_p^2}\frac{\Delta L}{L_0} & \text{if} & (n-1)(\frac{2L_0}{\Delta L} + 1) &<& \frac{W}{2mnv^2_p}&<& (n-1)(\frac{2L_0}{\Delta L} + 1) + \frac{2L_0}{\Delta L},\\
1 & \text{if} &(n-1)(\frac{2L_0}{\Delta L} + 1) + \frac{2L_0}{\Delta L} &<& \frac{W}{2mnv^2_p}&<& (n+1)(\frac{2L_0}{\Delta L} + 1)- \frac{2L_0}{\Delta L},\\
\hspace{0.25cm}(n+1)(1 + \frac{\Delta L}{2L_0}) - \frac{W}{4mnv_p^2}\frac{\Delta L}{L_0} & \text{if}& (n+1)(\frac{2L_0 }{\Delta L} + 1) - \frac{2L_0}{\Delta L} 
&<& \frac{W}{2mnv^2_p}&<& (n+1)(\frac{2L_0}{\Delta L} + 1).\\
\end{array} \right.
\end{equation} 
\end{widetext}
A variation of the piston velocity $v_p$ is equivalent to a variation of the temperature in the sense of the following scaling relation:
\begin{equation}\label{equ:PWequivalent}
	P(W,L_0,\frac{v_p}{\lambda},\beta) = \lambda^2 P(\lambda^2 W,L_0, v_p,\frac{\beta}{\lambda^2}).
\end{equation}
This equation, which can be easily derived directly from Eq. (\ref{equ:PW_gas}), holds, because due to the elastic hard collisions the same sequence of events occurs if both $v$ and $v_p$ are scaled by the same factor.

\begin{figure}[h]
\centerline{\includegraphics[width=7.0cm,clip=true]{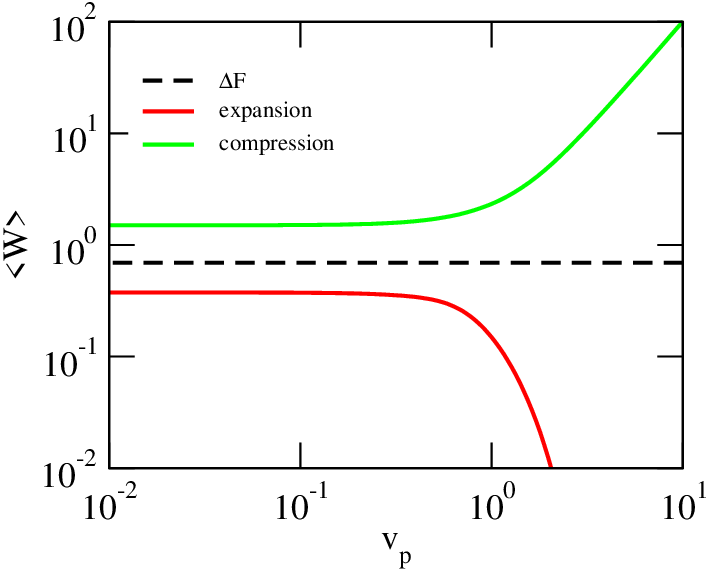}}
\caption{Average work $\langle W \rangle$ (solid red line) performed {\em by} the ideal gas during the expansion as a function of the piston velocity $v_p$ for $L_0=1$, $\Delta L=1$ and $\beta =1$. Also shown is the work done {\em on} the ideal gas during the compression (solid green line). The free energy difference $\Delta F=k_{\rm B}T\log [(L_0+\Delta L)/L_0]=\log(2)$ is shown as a dashed line. For $v_p \rightarrow 0$, the average work converges towards $(1/2)[1-L_0^2/(L_0+\Delta L)^2]=0.375$.}
\label{fig:figmeanworkjarzynski}
\end{figure}

The average work 
\begin{equation}\label{equ:meanworkint}
	\langle W \rangle = \int P(W) W dW 
\end{equation}
performed during the expansion of the cylinder from $L_0$ to $L=L_0+\Delta L$ calculated from the work distribution Eq. (\ref{equ:PW_gas}) by numerical integration is shown in Fig.  \ref{fig:figmeanworkjarzynski}. It is clear from the picture that the average work $\langle W \rangle$ is lower than the free energy difference $\Delta F=F(L_0)-F(L)=k_{\rm B}T\log [(L_0+\Delta L)/L_0]$ even in the limit of infinitely slow switching. In this limit, the process studied by Lua and Grosberg can be viewed as the quasistatic adiabatic expansion of an ideal gas where the average work follows from basic thermodynamic equations for the ideal gas:
\begin{eqnarray}
\label{meanworkideal}
\langle W \rangle & = & \int P dL = p_0 L_0^\gamma \int_{L_0}^L \frac{1}{L^\gamma} dL = \nonumber\\
&=& \frac{k_{\rm B}T}{1-\gamma}\left[\left(\frac{L}{L_0}\right)^{1-\gamma} -1 \right], 
\end{eqnarray}
where $\gamma=C_p/C_V=(f + 2)/f$ and $f$ is the number of degrees of freedom. In our one-dimensional case, $f=1$ and $\gamma=3$ such that
\begin{equation}
\label{equ:work_adiabatic}
\langle W \rangle  = \frac{k_{\rm B}T}{2} \left(\frac{L^2-L_0^2}{L^2}\right).
\end{equation}
For fast switching, on the other hand, the average work converges to zero because less and less collisions of the moving particle with the piston occur as the piston velocity is increased. Nevertheless, the Jarzynski equality holds over the whole range of piston velocities as can be directly verified from Eq. (\ref{equ:PW_gas}).

In the slow switching limit, i.e., for $v_p\ll \sqrt{(k_{\rm B}T/m)}\Delta L / (L_0+\Delta L)$, the work distribution can also be obtained by considering  adiabatic invariants \cite{EHRENFEST,LIBERMANN,HERTZ,GOLDSTEIN,IDGAS}. For a one-dimensional mechanical system the principle of adiabatic invariance states that the action variable 
\begin{equation}
I=\frac{1}{2\pi}\oint p\, dq
\end{equation}
is conserved under slow variation of an external parameter in the Hamiltonian. In the above equation the contour integral extends over one period of the motion for constant energy $E=p^2/2m$. For the Lua-Grosberg model this relation implies that the quantity $E L^2$ is conserved during a slow, i.e., adiabatic, transformation of the Hamiltonian. As a consequence, the energy $E$ of the system at the end of an expansion from $L_0$ to $L$ is a unique function of the initial energy $E_0$:
\begin{equation}
E=\left(\frac{L_0}{L}\right)^2E_0.
\end{equation}
Thus, the work carried out by the gas starting from any initial condition with energy $E_0$ is given by $W=E_0 - (L_0/L)^2E_0= E_0 [1-(L_0/L)^2]$. The work distribution in the slow switching limit can then be obtained as average over the canonically distributed initial conditions:
\begin{eqnarray}
\hspace{-0.6cm}P(W)& = & \nonumber \\
&&\hspace{-1cm}\sqrt{\frac{\beta}{2\pi m}}\int dp \;e^{-\beta p^2/2m}\delta\left[W-\frac{p^2}{2m}\left(1-\frac{L_0^2}{L^2}\right)\right].
\end{eqnarray}
Integration yields
\begin{equation}
P(W)=\sqrt{\frac{\beta L^2}{\pi(L^2-L_0^2)}}\frac{1}{\sqrt{W}}e^{-\beta W \left(\frac{L^2}{L^2-L_0^2}\right)},
\end{equation}
which is the one-particle one-dimensional special case of the work distribution derived by Crooks and Jarzynski for an ideal gas of $N$ particles \cite{IDGAS}. It is easy to verify directly that for this work distribution that the Jarzynski equality holds and the average work $\langle W\rangle$ performed by the gas is given by Eq. (\ref{equ:work_adiabatic}). Note that even tough the expansion occurs infinitely slowly in this case, the work distribution is not Gaussian and the width $\sigma_W$ and the average $\langle W \rangle$ of the work distribution are related by
\begin{equation}
\sigma_W=\frac{k_{\rm B}T}{\sqrt{2}} \left(\frac{L^2-L_0^2}{L^2}\right)=\sqrt{2}\langle W\rangle
\end{equation}
rather than by Eq. (\ref{equ:Wave}).

In the fast switching limit, i.e., for $v_p\gg \sqrt{(k_{\rm B}T/m)}\Delta L / (L_0+\Delta L)$, only trajectories with 0 or 1 collision yield important contributions to the work distribution. In this case, one obtains \cite{LUA_Illustration}:
\begin{eqnarray}
P(W) & = & P_0\delta (W)+\nonumber \\
& & \sqrt{\frac{m\beta}{2 \pi}}\left(\frac{\Delta L}{L_0} \right)\frac{W}{4v_p^3m^2}e^{-\frac{\beta}{8mv_p^2}(2mv_p^2+W)^2},
\end{eqnarray}
where $P_0$ is given by 
\begin{equation}
P_0=1-\frac{1}{\sqrt{2 \pi m\beta}}\left(\frac{\Delta L}{L_0} \right)\frac{1}{v_p}.
\end{equation}

In the following paragraphs we will also consider the statistical errors for the reverse process, namely the compression of the ideal gas from the cylinder length $L_0+\Delta L$ to $L_0$. The work distribution $P_R(W)$ for the reverse process is simply given by the Crooks relation Eq. (\ref{equ:crooks}), which, with the sign conventions for $W$ and $\Delta F$ used in this section reads:
\begin{equation}
\label{equ:PWR_gas}
P_R(-W)=P(W)e^{\beta (W-\Delta F)}=P(W)e^{\beta W}\frac{L_0}{L}.
\end{equation}

\subsubsection{Efficiency}

We next determine the efficiency of fast switching methods for the ideal gas expansion from $L_0$ to $L_0+\Delta L$ as well as for the compression from $L_0+\Delta L$ to $L_0$. In this case, we investigate the straight forward approach, biased sampling with $\exp(\beta W/2)$ umbrella function, and thermodynamic integration. We do not, however, consider the $1/P$ bias here, because the work distribution for the ideal gas expansion includes a $\delta$-peak that creates problems in this case. 

When estimating statistical errors of fast switching simulations another difficulty arises in some situations. As detailed in Sec. \ref{sec:error}, the mean square error $\epsilon^2_N$ as well as the bias $b_N$ can be calculated from integrals over the work distribution. In the case of the ideal gas expansion, however, some of the integrals from Eqs. (\ref{equ:integral_f})-(\ref{equ:integral_l}) diverge for certain expansion ratios. To see this, consider the mean square error estimated for a straightforward fast switching simulation using the sign convention of this section:
\begin{equation}
\epsilon^2_N=\frac{k_{\rm B}^2T^2}{N}\left(e^{-2\beta\Delta F}\langle e^{2\beta W}\rangle-1\right).
\end{equation}
The average $\langle \exp(2\beta W) \rangle=\int dW P(W) \exp(2\beta W) $ exists only if the work distribution $P(W)$ decays sufficiently rapidly with increasing W to compensate for the fast growth of $\exp(2\beta W)$. For large work values $W$, the work distribution for the ideal gas expansion behaves as
\begin{equation}
P(W) \sim \exp\left[-\beta W\left(\frac{L^2}{L^2-L_0^2}\right)\right],
\end{equation}
where we have neglected all factors that are immaterial for the convergence of the integral. Thus, $\langle \exp(2\beta W)\rangle$ is finite only if 
\begin{equation}
\frac{L^2}{L^2-L_0^2}-2 > 0,
\end{equation}
or, equivalently, if the ideal gas is expanded by no more than a factor of $\sqrt{2}$, $(L/L_0)< \sqrt{2}$. If this condition, which is independent of the switching rate, is not met, $X = \exp(\beta W)$ has infinite variance and the expressions from Sec.  \ref{sec:error} can not be applied to estimate the statistical error of the free energy estimate. In this case, a generalization of the law of large numbers \cite{PAPOULIS} still guarantees that the free energy converges to the correct value, but the fat-tailed work distribution prevents Gaussian statistics from arising even for large sample sizes.  

For umbrella sampling with the exponential bias $\exp(\beta W/2)$ a similar analysis of the average $\langle Y^2 \rangle_\pi$ reveals that the corresponding integral converges only provided that $(L/L_0)< \sqrt{3}$. For larger expansion ratios, the integral diverges for any switching rate. In contrast, the integrals associated with the thermodynamic integration scheme always converge regardless of the expansion ratio $L/L_0$ and the switching rate.

The situation is slightly different for the ideal gas compression. In this case, straightforward fast switching yields error estimates that are finite for any expansion ratio. Since for umbrella sampling with exponential bias  $\exp(\beta W/2)$ and thermodynamic integration the errors in forward and backward direction are identical, the same limitations as for the ideal gas expansion apply in these cases. 

In the following we will study the efficiency for expansion ratios of $L/L_0=1.2$ and $L/L_0=2.0$. For $L/L_0=1.2$ the error estimates for all fast switching methods are finite both for the expansion and the compression. For $L/L_0=2$ we consider only the thermodynamic integration method and straightforward fast switching for the ideal gas compression. All other methods yield diverging error estimates for this expansion ratio.   

\begin{figure}[h]
\centerline{\includegraphics[width=7.0cm,clip=true]{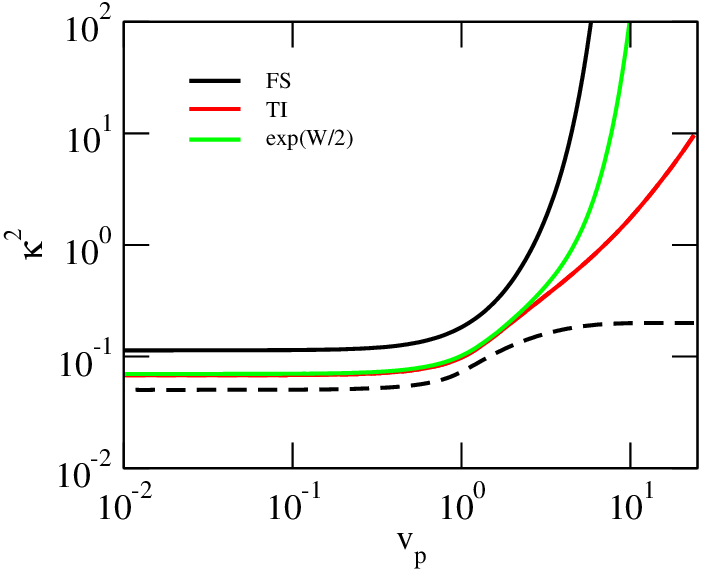}}
\caption{Error measure $\kappa^2=\epsilon_N^2 N$ as a function of the piston velocity $v_p$ for different fast switching methods and an expansion ratio of $L/L_0=1.2$. The black lines denote the straight forward fast switching errors for expansion (solid line) and compression (dashed line).}
\label{fig:figerrorgrosberg}
\end{figure}

Figure \ref{fig:figerrorgrosberg} shows the errors as a function of the piston velocity $v_p$ obtained by numerical integration of Eqs. (\ref{equ:integral_f})-(\ref{equ:integral_l}) for an expansion from $L_0 = 1$ to $L=L_0+\Delta L=1.2$. Here and in the subsequent calculations the inverse temperature of the initial state was $\beta = 1$. 
For infinitely slow and infinitely fast switching the errors can be calculated analytically and were found to agree perfectly with the results obtain by numerical integration. As can be seen in Fig. \ref{fig:figerrorgrosberg}, the errors become independent of the switching rate $v_p$ for all three methods in the slow switching limit. In the fast switching limit, the error grows rapidly with $v_p$ particularly for straightforward fast switching and the exponential bias. The thermodynamic integration scheme, for which the growth is slower, yields smaller errors for all switching rates. For all three cases the errors diverge for $v_p \to \infty$. The overall smallest errors, however, are obtained if the reverse process, i.e., the compression, is simulated with straightforward fast switching (dashed line in Fig. \ref{fig:figerrorgrosberg}). In the fast switching limit this error becomes $\Delta L / L_0$. Since the initial position of the particle is uniformly distributed in the interval $[0, L]$,  this value is simply the probability that the fast moving piston hits the particle at least once. Note that for thermodynamic integration and umbrella sampling with the exponential bias the errors are identical for expansion and compression.

\begin{figure}[h]
\centerline{\includegraphics[clip=true,width=7.0cm]{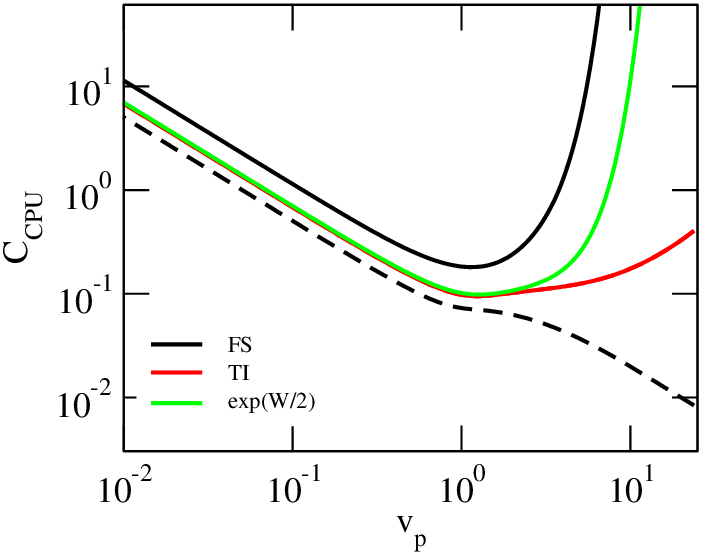}}
\caption{Computational cost $C_{\rm CPU}=\kappa^2\tau$ as a function of the switching rate $v_p$ for different fast switching methods and and expansion ratio of $L/L_0=1.2$. The black lines denote the straight forward fast switching errors for expansion (solid line) and compression (dashed line). }
\label{fig:figccpugrosberg}
\end{figure}

From the errors depicted in Fig. \ref{fig:figerrorgrosberg} one can determine the computational cost required to obtain an accuracy of $k_{\rm B}T$ in the free energy estimate [see Eq. (\ref{equ:CCPU})]. This computational cost is shown in Fig. \ref{fig:figccpugrosberg} for different fast switching methods and an expansion ratio of $L/L_0=1.2$ as a function of the switching rate. For straightforward fast switching in the expansion direction, thermodynamic integration and umbrella sampling with the exponential bias an optimum switching rate of about $v_p \sim 1$ exists. In the case of straightforward fast switching in the compression direction, however, no such minimum exists and the computational cost keeps decreasing for increasing switching rate. 

\begin{figure}[h]
\centerline{\includegraphics[width=7.0cm,clip=true]{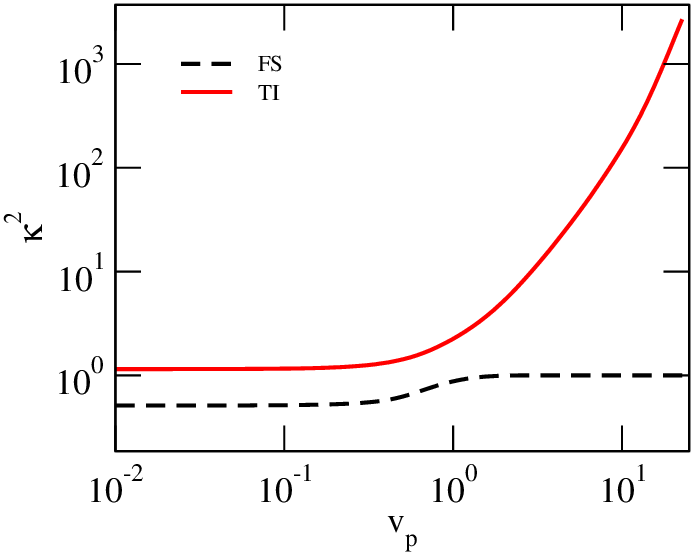}}
\caption{Error measure $\epsilon_N^2 N$ as a function of the piston velocity $v_p$ for the ideal gas expansion/compression with an expansion ration of $L/L_0=2.0$. The dashed black line denotes the straightforward fast switching errors (FS) and the solid red line the thermodynamic integration errors (TI).}
\label{fig:figerrorgrosbergL2}
\end{figure}

\begin{figure}[h]
\centerline{\includegraphics[clip=true,width=7.0cm]{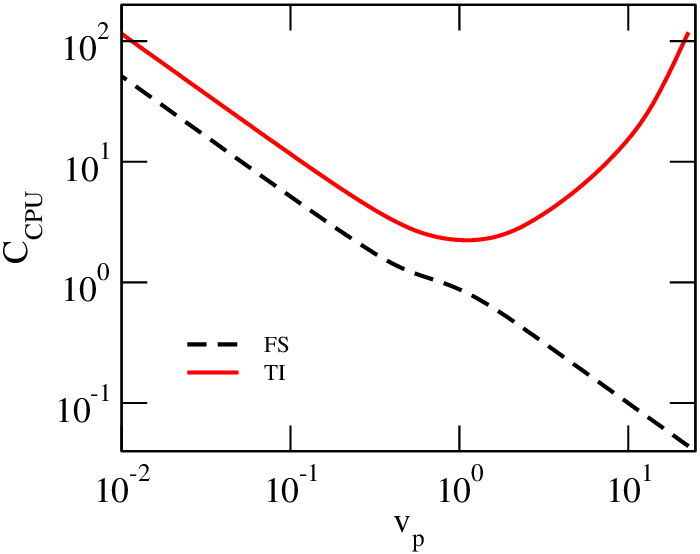}}
\caption{Computational cost $C_{\rm CPU}$ as a function of the switching rate $v_p$ for the ideal gas expansion/compression with an expansion ratio of $L/L_0=2$. The dashed black line denotes the straight forward fast switching results (FS) for the compression and the solid red line the thermodynamic integration results (TI). }
\label{fig:figccpugrosbergL2}
\end{figure}

As explained above, for an expansion ratio of $L/L_0=2$ our expressions yield finite statistical errors only for the thermodynamic integration method and for straightforward fast switching in the compression direction. The errors for these cases are shown in Fig. \ref{fig:figerrorgrosbergL2}. Straightforward fast switching of the ideal gas compression yields errors that are smaller than the thermodynamic integration errors for all piston velocities. Both in the slow and fast switching limit the straightforward fast switching errors become constant. Accordingly, the computational cost of the straightforward compression is decreasing as $1/v_p$ in the fast switching limit while the thermodynamic integration error goes through a minimum at about $v_p \approx 1$ and then rapidly grows for increasing piston velocity $v_p$ (see Fig. \ref{fig:figccpugrosbergL2}). Thus both for $L/L_0=1.2$ and $L/L_0=2.0$ the straightforward fast switching simulation of the ideal gas compression is the most efficient among the methods studied here. However, optimum efficiency is obtained for infinitely fast compression, in which case  straightforward fast switching reduces to Zwanzig's perturbative approach applied to configurations rather than trajectories \cite{Zwanzig}.   

\subsubsection{Many-particle gas}

The model discussed in the previous section consisted of a single particle in one dimension. In this section, we study the expansion of an ideal gas of $M$ particles and analyze how the error depends on the number of particles. The system of strictly non-interacting particles considered here is similar, but not identical to the dilute gas of weakly interacting particles studied by Jarzynski and Crooks in Ref. \cite{CROOKS_JAR_DILUTE}. For the $M$-particle ideal gas, the work performed on the piston during the expansion is the sum of the contributions of each particle.
\begin{equation}
\label{equ:work_assum}
W=\sum_{i=1}^{M} W_i.
\end{equation}
Here $W_i$ is the work performed on the system by particle $i$. According to Eq. (\ref{equ:epsilon}), the statistical error in the free energy calculated with straightforward fast switching from $N$ repetitions of the process is given by 
\begin{equation}
\epsilon_N^2 =\frac{k_{\rm B}^2T^2}{N}\left[ \frac{\langle X^2 \rangle}{\langle X\rangle^2}-1\right].
\end{equation}
For the $M$-particle system, the averages $\langle X \rangle$ and $\langle X^2 \rangle$ are 
\begin{equation}
\langle X \rangle=\langle e^{-\beta \sum_i W_i}\rangle =\langle X_1 \rangle^M  
\end{equation}
and 
\begin{equation}
\langle X^2 \rangle=\langle e^{-\beta \sum_i 2W_i}\rangle =\langle X^2_1 \rangle^M,  
\end{equation}
where $\langle X_1 \rangle$ and $\langle X^2_1 \rangle$ are the corresponding averages of the one-particle system. Here, we have used the fact that work contributions of the single particles are identical and statistically independent. The error $\epsilon_N(M)$ for the $M$-particle system becomes 
\begin{equation}
\label{equ:errormulti}
\epsilon^2_N(M)  = \frac{k_B^2 T^2}{N} \left[ \left( \frac{\langle X_1^2 \rangle}{\langle X_1 \rangle^2} \right)^M - 1 \right].
\end{equation}
The fraction $\langle X_1^2 \rangle/\langle X_1 \rangle^2$ is always larger than 1 and therefore the error $\epsilon$ diverges exponentially with the particle number in the case of straightforward fast switching. 

A similar analysis can be carried out for the error of work biased umbrella sampling simulations. For an umbrella function $\pi(W)$ that factorizes into a product of single particle umbrella functions $\pi_1(W_i)$ (this is for instance the case for the exponential bias $\exp(-\beta W/2)$), 
\begin{equation}
\pi(W)=\prod_{i=1}^{M}\pi_1(W_i),
\end{equation} 
the error is given by: 
\begin{eqnarray}
\label{equ:errormultiworkbiased}
\epsilon^2_N(M)  &=& \frac{k_B^2 T^2}{N}  \left[\left( \frac{\langle X_1^2 \rangle_{\pi}}{\langle X_1 \rangle^2_{\pi}} \right)^M    +   \left( \frac{\langle Y_1^2 \rangle_{\pi}}{\langle Y_1 \rangle^2_{\pi}} \right)^M\right.  \nonumber   \\
 & &  \left. -2 \left( \frac{\langle X_1 Y_1 \rangle_{\pi}}{\langle X_1 \rangle_{\pi} \langle Y_1 \rangle_{\pi}} \right)^M \right].
\end{eqnarray}
Here, as in the previous paragraph, the averages are single-particle averages. For large $M$, this expression will be dominated by either $(\langle X_1^2 \rangle_{\pi}/\langle X_1 \rangle^2_{\pi})^M$ or $(\langle Y_1^2 \rangle_{\pi}/\langle Y_1 \rangle^2_{\pi})^M$ depending on whether $\langle X_1^2 \rangle_{\pi}/\langle X_1 \rangle^2_{\pi}$ or $\langle Y_1^2 \rangle_{\pi}/\langle Y_1 \rangle^2_{\pi}$ is larger. (It is easy to see that the larger one of $\langle X_1^2 \rangle_{\pi}/\langle X_1 \rangle^2_{\pi}$ and $\langle Y_1^2 \rangle_{\pi}/\langle Y_1 \rangle^2_{\pi}$ is also larger than $\langle X_1Y_1 \rangle_\pi/ \langle X_1 \rangle_\pi \langle Y_1 \rangle_\pi$, which is always positive.) Thus, also in this case, the error in the free energy grows exponentially with the system size $M$.

For the work biased thermodynamic integration the error results from an integral over the work variance at different values of an auxiliary parameter $\alpha$ (see Eq. (\ref{equ:var_dFN})). Since in the $M$-particle ideal gas the work $W$ is a sum of $M$ independent contributions, the work variance of the entire system is $M$ times the single particle work variance for each value of $\alpha$. Hence, the $M$-particle error $\epsilon(M)$ scales with the square root of the number of particles, 
\begin{equation}
\label{equ:errormultiti}
\epsilon^2_N(M) = M \epsilon^2_N(1),
\end{equation}
where $\epsilon_N(1)$ is the error for the one-particle gas. Due to this kind of scaling, the thermodynamic integration method is superior to all other fast switching methods studied here for a strictly non-interacting ideal gas.

\section{Conclusion}
\label{sec:conclusion}

For irreversible transformations, the exponential average of Eq. (\ref{equ:jarzynski}) is dominated by rare, but important work values causing possibly large statistical inaccuracies in fast switching free energy computations. Several path sampling methods have proposed that aim at reducing these inaccuracies by enhanced sampling of those trajectories that contribute most to the average. In this paper, we investigate how these methods perform in the case of two simple models for which the work distribution is known analytically. 

For Gaussian work distributions, the computational effort required to achieve a given accuracy in the free energy can be determined analytically. In the case of a particle pulled through a viscous fluid, flat histogram sampling as well as Sun's work biased thermodynamic integration perform best. For both methods optimum efficiency is obtained in the limit of infinitely fast switching in which case these path sampling methods reduce to conventional flat histogram sampling and thermodynamic integration in configuration space rather than in trajectory space. 

The errors obtained in the case of the ideal gas expansion/compression, for which the work distribution was determined analytically by Lua and Grosberg \cite{GROSBERG_LUA_PEDAGOGICAL, LUA_Illustration}, show a qualitatively different behavior. In contrast to the particle pulled through a fluid, the work distribution for the expansion/compression process does not approach a constant shape in the fast switching limit. Rather, the work distribution keeps changing even for very large piston velocities. As a consequence, the computational cost of all methods goes through a minimum for intermediate piston velocities and grows for rapid expansion. The only exception here is the straightforward fast switching simulation of the gas compression, which is superior to the path sampling methods for all piston speeds. In this case, the computational cost keeps decreasing even for very large piston speeds where the fast switching simulation reduces to Zwanzig's perturbative approach. It is interesting that for certain expansion ratios only thermodynamic integration and straightforward fast switching of the compression process yield finite error estimates, while all other method yield diverging errors. This behavior is due to the fat tails of the work distribution that occur for larger expansion ratios.  

The superiority of straightforward fast switching for the ideal gas compression is most likely peculiar to the one-particle case. In the many-particle gas, the errors scale exponentially with particle number for all methods except for thermodynamic integration, in which case they scale with the square root of the particle number. Thus, for an ideal gas consisting of many particles Sun's work biased thermodynamic integration method seems to offer the most efficient free energy calculations. Since for this method the efficiency is maximum at intermediate piston velocities, this is an example where a fast switching method for the calculation of free energies is more efficient than conventional configuration space methods.

In general, the efficiency of a fast switching simulation depends on whether the process is carried out in forward or reverse direction. In particular, a straightforward fast switching simulation is more efficient in the direction in which more work is dissipated \cite{JARZYNSKI_RAREEVENTS} as confirmed here for the Lua-Grosberg ideal gas expansion and compression. For all the path sampling methods studied here, however, it can be shown using the Crooks identity that the work distribution is effectively symmetrized in a way that makes the corresponding efficiency identical in both directions. 

In summary, path sampling methods such as work biased thermodynamic integration or work biased umbrella sampling may help to reduce the statistical errors encountered in 
the calculation of equilibrium free energies from non-equilibrium simulations. The examples studied in this paper indicate that often conventional free energy estimation methods are likely to be superior to fast switching simulation. Nevertheless, advanced fast switching methods such as Sun's work biased thermodynamic integration may be competitive particularly if a large number of degrees of freedom significantly contribute to the free energy change that one wants to compute. 

\begin{acknowledgments}
This work was initiated during a Junior Fellowship of W.L. at the International Erwin Schr\"odinger Institute for Mathematical Physics (ESI) and completed with support from the Austrian Science Fund (FWF) under Grant No. P17178-N02. The authors acknowledge useful discussions with Phillip Geissler and Elisabeth Sch\"oll-Paschinger. 
\end{acknowledgments}

\bibliographystyle{prsty}

\begin{thebibliography}{10}

\bibitem{FRENKEL_SMIT} D. Frenkel and B. Smit, ``Understanding Molecular Simulation'', Academic Press, San Diego (2002).

\bibitem{FREE_ENERGY_BOOK} C. Chipot and A. Pohorille (Eds.), ``Free energy calculations. Theory and applications in chemistry and biology'', Springer Verlag, Berlin (2006).  

\bibitem{jarz} C.~Jarzynski, {\em Phys. Rev. Lett.} {\bf 78}, 2690 (1997).

\bibitem{JARZ_PRE_97} C. Jarzynski, {\em Phys. Rev. E} {\bf 56}, 5018 (1997).

\bibitem{gavin_jstatphys} G. E. Crooks, {\em J. Stat. Phys.}
{\bf 90}, 1481 (1998).

\bibitem{Hummer_Szabo}
G. Hummer and A. Szabo, {\em Proc. Natl. Acad. Sci. USA} {\bf 98}, 3658 (2001).

\bibitem{Evans}
D.J. Evans, Mol. Phys. {\bf 101}, 1551 (2003).

\bibitem{Dellago_Paschinger} E. Sch\"oll-Paschinger and C. Dellago, {\em J. Chem. Phys.},  {\bf 125}, 054105 (2006).

\bibitem{Kirkwood} J. G. Kirkwood, {\em J. Chem. Phys.} {\bf 3}, 300 (1935).

\bibitem{Zwanzig} R. Zwanzig, {\em J. Chem. Phys.} {\bf 22}, 1420 (1954).

\bibitem{jarTPS} H. Oberhofer, C. Dellago, P. L. Geissler,
{\em J. Phys. Chem. B} {\bf 109}, 6902 (2005).

\bibitem{SUN} Sean X. Sun,
{\em J. Chem. Phys.} {\bf 118}, 5769 (2003).

\bibitem{YTREBERG} F. M. Ytreberg and D. M. Zuckerman, {\em J. Chem. Phys.}{\bf 120}, 10876 (2004).

\bibitem{LECHNER} W. Lechner, H. Oberhofer, C. Dellago, and P. L. Geissler, {\em J. Chem. Phys.} {\bf 124}, 044113 (2006).

\bibitem{MAZONKA_JARZYNSKI} O. Mazonka and C. Jarzynski, {\em arxiv:cond-mat/9912121v1} (1999).

\bibitem{GROSBERG_LUA_PEDAGOGICAL} A. Grosberg and Rh. Lua, {\em J. Phys. Chem. B} {\bf 109} (14), 6805 (2005). 

\bibitem{BUSTAMANTE} J. Gore, F. Ritort, and C. Bustamante, {\em Proc. Natl. Acad. Sci.} {\bf 100}, 12564 (2003).     

\bibitem{JARZYNSKI_RAREEVENTS} C. Jarzynski, {\em Phys. Rev. E} {\bf 73}, 046105 (2006).

\bibitem{WIDOM} B. Widom, {\em J. Chem. Phys.} {\bf 39}, 2808 (1963).

\bibitem{TPS1} C. Dellago, P. G. Bolhuis, F. S. Csajka, and D. Chandler,
   {\em J. Chem. Phys. }{\bf  108}, 1964 (1998).

\bibitem{TPS2} C. Dellago, P. G. Bolhuis, P. L. Geissler, {\em Adv. Chem. Phys.} {\bf 123}, 1 (2002).


\bibitem{ATHENES}
M. Ath\`enes, {\em Eur. Phys. J. B} {\bf 38}, 651 (2004).

\bibitem{WANG_LANDAU} F. Wang and D. P. Landau, {\em Phys. Rev.
Lett.} {\bf 86}, 2050 (2001).

\bibitem{BERG} B. Berg and T. Neuhaus, {\em Phys. Lett. B} {\bf 267},
249 (1991);
B. Berg and T. Neuhaus, {\em Phys. Rev. Lett.} {\bf 69}, 9 (1992).

\bibitem{GEYER} C. J. Geyer and E. A. Thompson, {\em J. Am. Stat.
Soc.} {\bf 90}, 909 (1995).

\bibitem{ZUCKERMAN_WOOLF} D. Zuckerman and T. Woolf, {\em Phys. Rev. Lett.} {\bf 89}, 180602 (2002). 

\bibitem{CROOKS} G.E. Crooks, {\em Phys. Rev. E} {\bf 60}, 2721 (1999). 

\bibitem{BAULE} A. Baule and R.M.L. Evans and P.D. Olmsted, {\em arXiv:cond-mat/0607575} (2006).

\bibitem{EHRENFEST} P. Ehrenfest, {\em Ann. Phys. (Leipzig)} {\bf 51}, 327 (1916).

\bibitem{LIBERMANN} A. J. Lichtenberg and M. A. Libermann, {\em
Regular and Stochastic Motion}, Springer-Verlag, New York (1983).

\bibitem{HERTZ} P. Hertz, {\em Ann. Phys. (Leipzig)} {\bf 33}, 225 (1910); {\em ibid.}{\bf 33}, 537 (1910) .

\bibitem{GOLDSTEIN} H. Goldstein, {\em Classical Mechanics}, Addison-Wesley (1987).

\bibitem{IDGAS} G. E. Crooks and C. Jarzynski, cond-mat/0603116 (2006).

\bibitem{LUA_Illustration} Rh. Lua, {\em arxiv:cond-mat/00511302} (2005).

\bibitem{PAPOULIS} A. Papoulis, {\em Probability, random variables, and stochastic processes},  McGraw-Hill Education (1981).

\bibitem{CROOKS_JAR_DILUTE} G. Crooks and C. Jarzynski, {\em arxiv:cond-mat/0603116v1} (2006).

\end{thebibliography}

\end{document}